\documentclass[twocolumn,aps,showpacs,prb,tightenlines,amsmath,amssymb,superscriptaddress]{revtex4}
\usepackage{bm}
\usepackage{graphicx}
\usepackage{bmpsize}             
\usepackage{amssymb}                                                             
\usepackage{amsmath}
\usepackage{colordvi}

\newcommand{\bgreek}[1]{\mbox{\boldmath$#1$\unboldmath}}
\begin{document} 

\title{Spin-Orbit Qubit on a Multiferroic Insulator in a Superconducting Resonator}

\author{P. Zhang}
\affiliation{Center for Emergent Matter Science, RIKEN, Saitama 351-0198, Japan}
\author{Ze-Liang Xiang}
\affiliation{Center for  Emergent Matter Science, RIKEN, Saitama 351-0198, Japan}
\author{Franco Nori}
\affiliation{Center for Emergent Matter Science, RIKEN, Saitama 351-0198, Japan}
\affiliation{Department of Physics, The University of Michigan, Ann Arbor, Michigan 48109-1040, USA}
\date{\today}

\begin{abstract} 
We propose a spin-orbit qubit in a nanowire quantum dot on the
surface of a multiferroic insulator with a cycloidal spiral magnetic order. The
spiral exchange field from the multiferroic insulator causes an inhomogeneous Zeeman-like interaction on the electron spin in the
quantum dot, producing a spin-orbit qubit. The absence of an
external magnetic field benefits the integration of such spin-orbit qubit into
high-quality superconducting resonators. By exploiting the Rashba spin-orbit coupling in
the quantum dot via a gate voltage, one can obtain an effective
spin-photon coupling with an efficient on/off switching. This makes the proposed
device controllable and promising for hybrid quantum communications.
\end{abstract}
\maketitle
\section{Introduction}
Spin-based qubits, owing to their long coherence times and individual coherent manipulation, are promising candidates for building blocks
of quantum information processors.\cite{wuReview,buluta104401,loss120,hanson1217,zwanenburg961} A conventional
spin qubit can be simply realized via Zeeman splitting of two Kramers-degenerate states by a static magnetic field and controlled by an ac magnetic
field.\cite{hanson1217,awschalom1174,koppens766,engel4648} However, its application is
limited due to the difficulty in generating and localizing an ac magnetic field at
the nanoscale. Owing to the interplay between spin and orbital degrees of freedom,
the spin-orbit qubit allows the possibility for manipulating spins via an easily-accessible ac electric field, i.e., by means of the
electric-dipole spin resonance (EDSR).\cite{rashba126405,rashba137}
Intuitively, the interplay between spin and orbit can arise from the spin-orbit coupling (SOC),
e.g., the Rashba or Dresselhaus type, which
couples the electron spin ${\bgreek \sigma}$ to the momentum ${\bf p}$. The SOC-mediated EDSR has been
widely studied in the literature.\cite{Nowack1430,perge166801,li086805,sadreev115302,golovach165319,hu035314,khomitsky125312,Arrondo155328}
Instead of invoking SOC, an alternative way to achieve the interplay between spin and
orbit is coupling the electron spin ${\bgreek \sigma}$ to the
coordinate ${\bf r}$. This spin-coordinate coupling can be accomplished by,
e.g., an inhomogeneous Zeeman-like interaction \cite{tokura047202,kato1201,ladriere776,obata085317,hu035314,cottet160502} or a fluctuating
hyperfine interaction.\cite{laird246601,shafiei107601}

Apart from coherent manipulation, scaling up the spin-orbit-qubit
architecture also involves quantum information storing and transferring. Embedding the spin-orbit qubit into
a cavity resonator to achieve spin-photon coupling seems particularly attractive, as
the mobile photons in the cavity can store and transfer quantum information with little loss
of coherence.\cite{xiang623} Indeed, in view of their energy scales, the semiconductor-based spin-orbit qubit is compatible with
the superconducting microwave resonator. Moreover, integrating the spin-orbit qubit into the superconducting cavity promotes hybrid quantum
communications, e.g., in combination with superconducting
qubits or charge qubits.\cite{xiang623,you42,lambert17005} Several proposals for coupling
spin-orbit qubits to superconducting cavities have been reported.\cite{burkard041307,jin190506,cottet160502,hu035314} However, the 
spin-orbit qubit invoking SOC requires an external static magnetic field,\cite{Nowack1430,li086805,perge166801,Arrondo155328} which 
is not naturally compatible with superconducting cavities of high quality
factors.\cite{cottet160502} Therefore, a spin-orbit qubit
without an external magnetic field is preferred for constructing a hybrid
system. It has been proposed that, by using an inhomogeneous Zeeman-like
interaction induced by ferromagnetic contacts or micromagnets,\cite{tokura047202,obata085317,ladriere776,cottet160502} one can
realize spin-orbit qubits in the absence of a magnetic field and effectively
couple them to superconducting cavities.\cite{cottet160502,hu035314}

In this work, we propose a spin-orbit qubit mediated by the spin-coordinate
coupling and study its coupling to a superconducting coplanar waveguide
resonator. Different from previous studies,\cite{tokura047202,kato1201,laird246601,shafiei107601,ladriere776,obata085317,hu035314,cottet160502} our
proposal relies on the inhomogeneous exchange field arising from the
multiferroic insulators with a cycloidal spiral magnetic
order.\cite{kimura387,lebeugle024116,rovillain975,khomskii20,thomas423201,yamasaki147204} 
These multiferroic insulators provide a unique opportunity for the design of
functional devices owing to the cycloidal spiral magnetic order as well as the magnetoelectric coupling.\cite{zhang014433,zhai022107,thomas423201} In
our setup for the spin-orbit qubit, as illustrated in Fig.~\ref{fig1}(a), a
gated nanowire with a quantum dot is placed on top of a multiferroic 
insulator. The spiral exchange field arising from the magnetic moments in the
multiferroic insulator causes an inhomogeneous Zeeman-like interaction on the
quantum-dot spin. Therefore, a spin-orbit qubit is produced in the nanowire quantum
dot. The absence of an external magnetic field facilitates the
integration of the spin-orbit qubit into the superconducting coplanar
waveguide, as illustrated in Fig.~\ref{fig1}(b). In
this hybrid circuit, both the level spacing of the spin-orbit qubit and the
spin-photon coupling depend on the ratio between the dot size
and the wavelength of the spiral magnetic order in the substrate. When the Rashba SOC is introduced into
the nanowire, the level spacing and spin-photon coupling can be
adjusted by tuning the Rashba SOC via a gate voltage on the
nanowire. With the modulation of the Rashba SOC, we can obtain an effective
spin-photon coupling with an efficient on/off switching. This is promising for manipulating, storing and transferring quantum
information in the data bus provided by the circuit cavity. 

This paper is organized as follows. First, we establish the
spin-orbit qubit on the surface of a multiferroic insulator. After that, we integrate the spin-orbit
qubit into a superconducting coplanar waveguide and study its spin-photon
coupling. We further study the modulation of the
Rashba SOC on the hybrid system. At last we discuss the
experimental realizability of the proposed device. 

\section{Spin-orbit qubit on a multiferroic insulator}

\begin{figure}[thb]
{\includegraphics[width=8cm]{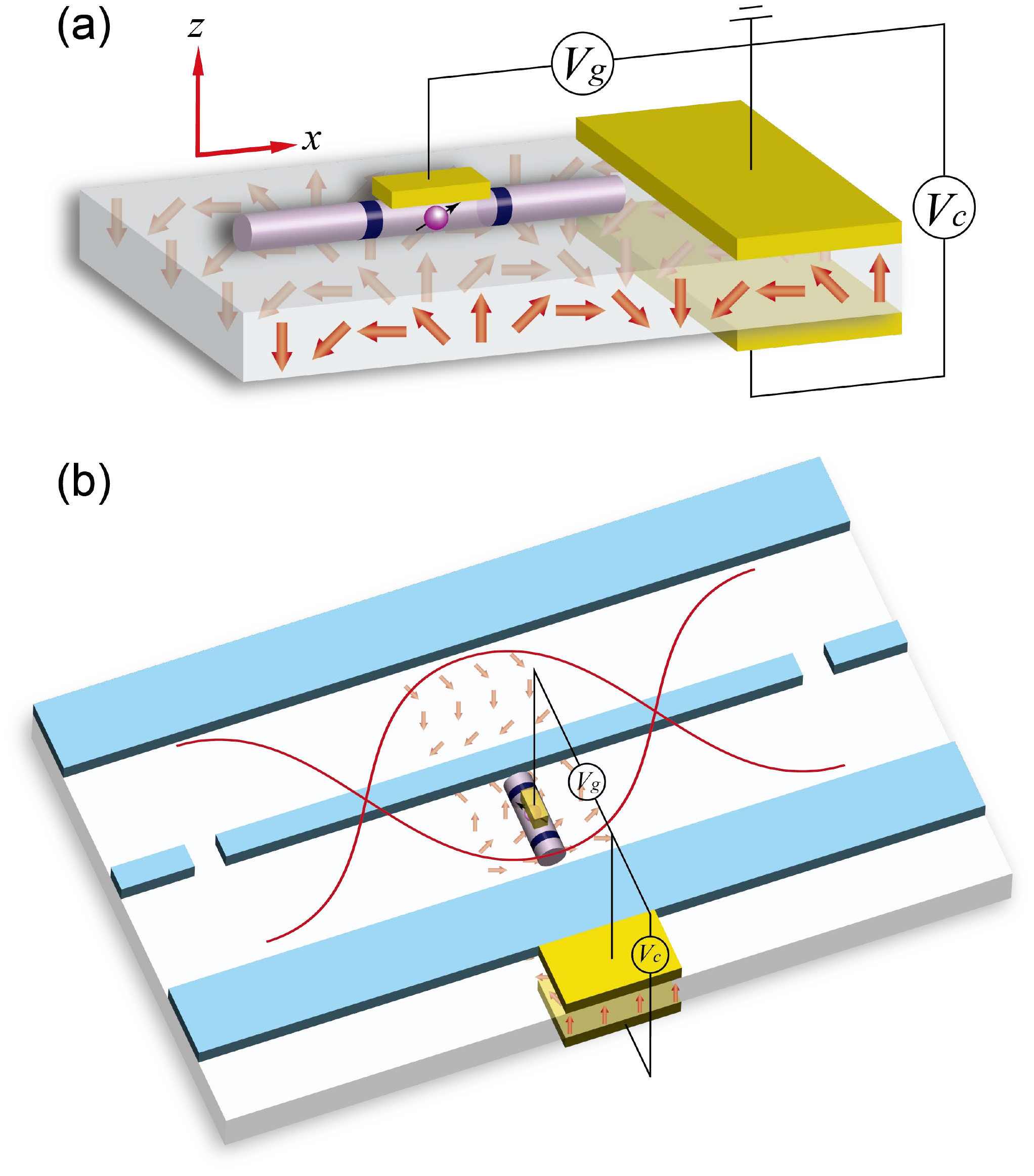}}
    \caption{(Color online) (a)~Schematic of the proposed spin-orbit qubit: a
      gated nanowire on the surface of a multiferroic insulator. The nanowire is aligned
      along the propagation direction of the spiral magnetic order in the
      multiferroic insulator, indicated by the series of rotating arrows. Two
      gate electrodes (indicated by the two dark blue ring-shaped contacts) supply a parabolic confining
      potential and form a quantum dot in between. The gate electrode on the top of the quantum dot with voltage
      $V_g$ controls the Rashba SOC. Two gate electrodes on the top
      and bottom of the multiferroic insulator supply a voltage $V_c$ which controls the spiral helicity of
      the magnetic order. (b)~Schematic of the integration of the spin-orbit
      qubit into a superconducting coplanar waveguide resonator. The nanowire
      is placed parallel to the electric field between the center conductor and
      the ground plane and located at the maximum of the electric field. Note
      that in the multiferroic-insulator substrate, only the magnetic moments near the nanowire are schematically
      shown by the rotating arrows.}
   \label{fig1}
 \end{figure}

Our study starts from the device schematically shown in Fig.~\ref{fig1}(a). In this
setup, a nanowire lies on the surface of a multiferroic insulator,
e.g., TbMnO$_3$ or BiFeO$_3$,\cite{lebeugle024116,rovillain975,khomskii20,yamasaki147204,thomas423201}
and is aligned parallel to the propagation direction of the spiral magnetic moments in
the substrate. The nanowire is gated by
two electrodes producing a quantum dot, which is assumed to be subject to a 1D
parabolic potential.

We consider a single electron in the quantum dot. In the coordinate system
with the $x$-axis along the nanowire and the $z$-axis
perpendicular to the top surface of the multiferroic insulator, the electron is described by the Hamiltonian,
\begin{align}
H=\frac{p^2}{2m_e}+\frac{1}{2}m_e\omega^2x^2+{\bf J}(x)\cdot{\bgreek \sigma}.
\label{hamil}
\end{align}
Here $m_e$ is the effective electron mass and $p=-i\hbar\partial_x$ is the momentum
operator. The second term in the Hamiltonian is the parabolic potential. The
last term depicts the interaction between the electron spin ${\bgreek \sigma}$ and the exchange
field from the cycloidal spiral magnetic moments, 
\begin{align}
{\bf J}(x)=J(\sin(\chi qx+\phi),0,\cos(\chi q x+\phi)).
\end{align}
In writing this term we have assumed the dot size $x_0=\sqrt{\hbar/(m_e\omega)}$ to be much larger than the distance
($\sim$~0.1~nm) between the nearby magnetic atoms in the substrate. Here $q=2\pi/\lambda$ is
the wavevector of the spiral order corresponding to a wavelength $\lambda$ and $\phi$ ($0\le \phi<2\pi$) is the phase of the
exchange field at $x=0$. The spiral helicity $\chi$ ($=\pm 1$) of the magnetic order is reversable
by a gate voltage on the multiferroic insulator [as illustrated by $V_c$ in Fig.~\ref{fig1}(a)] due to the
magnetoelectric coupling.\cite{yamasaki147204} The strength of the exchange coupling $J$ (we assume $J>0$) between the electron spin and the magnetic
moments, depending on their distance and the specific hosts, is weak and assumed to be of the order
of 1-10~$\mu$eV.\cite{cottet160502,tokura047202} 

Due to the spiral geometry of the magnetic order, the macroscopic magnetism of the multiferroic
insulator is zero, while the exchange coupling still breaks the time-reversal
symmetry locally and causes an inhomogeneous Zeeman-like interaction on the
quantum-dot spin. In the presence to this inhomogeneous Zeeman-like
interaction, a spin-orbit qubit is realizable in the quantum dot. One can also understand the availability of a spin-orbit qubit
in our setup in the spiral frame with the spin ${z}$-axis along the local
magnetic moment. Using a unitary transformation ${\tilde H}=U^\dagger(x)HU(x)$, where
$U(x)=\exp{[-i(\chi q x+\phi)\sigma_y/2]}$,\cite{zhang014433} one arrives at
\begin{align}
   {\tilde H}=\frac{p^2}{2m_e}+\frac{1}{2}m_e\omega^2x^2-\alpha_0 p\sigma_y+J\sigma_z+\frac{\hbar^2q^2}{8m_e},\label{rl}
\end{align}
where $\alpha_0=\chi\hbar q/(2m_e)$. This Hamiltonian evidently indicates that
in the spiral frame, the exchange field supplies not only the homogeneous Zeeman-like
interaction, $J\sigma_z$, but also an effective Rashba-like SOC, $-\alpha_0 p\sigma_y$.\cite{zhang014433} This Hamiltonian is equivalent to the one studied in
Ref.~\onlinecite{li086805}, where a spin-orbit qubit was realized by virtue of an external magnetic
field and the genuine Rashba/Dresselhaus SOC.

Now we demonstrate the realization of a spin-orbit qubit by studying the
low-energy bound states in the quantum dot. For the reasonable case with
$J/(\hbar\omega)\lesssim 0.1$, the exchange
coupling can be treated as a perturbation. We rewrite the Hamiltonian (\ref{hamil}) as
$H=H_0+H_1$, where $H_1={\bf J}(x)\cdot{\bgreek\sigma}$. The eigenstates of $H_0$, describing a
harmonic oscillator, can be written as $|n\pm\rangle=|n\rangle|\pm\rangle$ with the eigenenergies
$\varepsilon_n=\left(n+\frac{1}{2}\right)\hbar\omega$ ($n=0,1,2...$). Here $|n\rangle$ is
the orbital eigenstate of the harmonic oscillator and $|+\rangle$
($|-\rangle$) is the spin-up (-down) eigenstate of $\sigma_z$. We focus on
the $n=0$ Hilbert subspace which is two-fold degenerate. First-order degenerate perturbation theory gives the lowest
two bound states of $H$ with energies $\varepsilon_{0\pm}=\varepsilon_0\pm
\hbar\Delta/2$, where
\begin{align}
\Delta=\Delta_0\exp{(-\eta^2)},
\end{align}
with $\Delta_0=2J/\hbar$ and $\eta=\chi\pi x_0/\lambda$. The corresponding wavefunctions are
\begin{align}\nonumber
  |\widetilde{0\pm}\rangle=&e^{-i\phi\sigma_y/2}\Big\{|0\pm\rangle-\frac{J
    e^{-\eta^2}}{\hbar\omega}\sum_{m=1}^{+\infty}\Big[\frac{\pm(i\sqrt{2}\eta)^{2m}}{2m\sqrt{(2m)!}}|2m\pm\rangle
  \\&-\frac{i(i\sqrt{2}\eta)^{2m-1}}{(2m-1)\sqrt{(2m-1)!}}|2m-1\mp\rangle\Big]\Big\}.
\end{align}
The two lowest bound states $|\widetilde{0\pm}\rangle$, spaced by $\hbar\Delta$ and about $\hbar\omega$ away from the
nearest higher-energy state, can be used to encode the spin-orbit
qubit. As a result, with the aid of the spiral exchange field supplied by a
multiferroic insulator, we realize a spin-orbit qubit in the absence of an
external magnetic field as well as the Rashba/Dresselhaus SOC. 

\section{Spin-photon coupling in a superconducting cavity}
The spin-orbit qubit can respond to an ac electric field, via EDSR.\cite{rashba126405,rashba137} Due to the small level spacing, the spin-orbit
qubit is controllable by low-temperature microwave technology. This can be accomplished by virtue of a superconducting
resonator, which works at temperatures $\sim$~mK with resonance frequencies
$\sim$~GHz.\cite{xiang623} Indeed, integrating spin-orbit qubits into
superconducting resonators has recently attracted much
interest,\cite{burkard041307,cottet160502,jin190506,hu035314}
to explore novel hybrid quantum circuits.\cite{xiang623} Moreover, the spin-orbit qubit proposed here,
which is external-magnetic-field-free, is naturally compatible with
superconducting resonators of high quality factors.

As schematically shown in Fig.~\ref{fig1}(b), we embed the spin-orbit qubit into a superconducting coplanar
waveguide,\cite{xiang623,hu035314} with the nanowire aligned parallel to the
electric field between the center conductor and the ground plane. The resonant
photon energy ($\sim$~GHz) is too low to excite magnons in the
multiferroic-insulator substrate,\cite{rovillain975} and we assume that the spiral
magnetic order keeps steady during the operation
of the spin-orbit qubit. The spin-orbit qubit, photons, as well as their coupling, can be described by the 
Hamiltonian,\cite{xiang623}
\begin{align}
H_{\rm eff}=\frac{\hbar\Delta}{2}s_z+\hbar\omega_r\left(a^\dagger a+\frac{1}{2}\right)+\hbar g(a^\dagger s_-+a s_+).
\label{edsr}
\end{align}
Here $a$ ($a^\dagger$) is the annihilation (creation) operator for photons with
frequency $\omega_r$ in the cavity, and $s_{x,y,z}$ are the Pauli matrices in
the $|\widetilde{0\pm}\rangle$ subspace with $s_\pm=(s_x\pm i s_y)/2$. The
spin-photon coupling strength
\begin{align} 
g=\langle\widetilde{0+}|x|\widetilde{0-}\rangle Ee/\hbar,
\end{align}
where $E$ is the cavity electric field on the spin-orbit qubit. Up to first order in
$J/(\hbar\omega)$,
\begin{align}
g=g_0(x_0/\lambda)^3\eta \exp{(-\eta^2)},
\end{align}
where $g_0=-eEm_eJ\lambda^3/\hbar^3$. 

\begin{figure}[hbt]
  {\includegraphics[width=9cm]{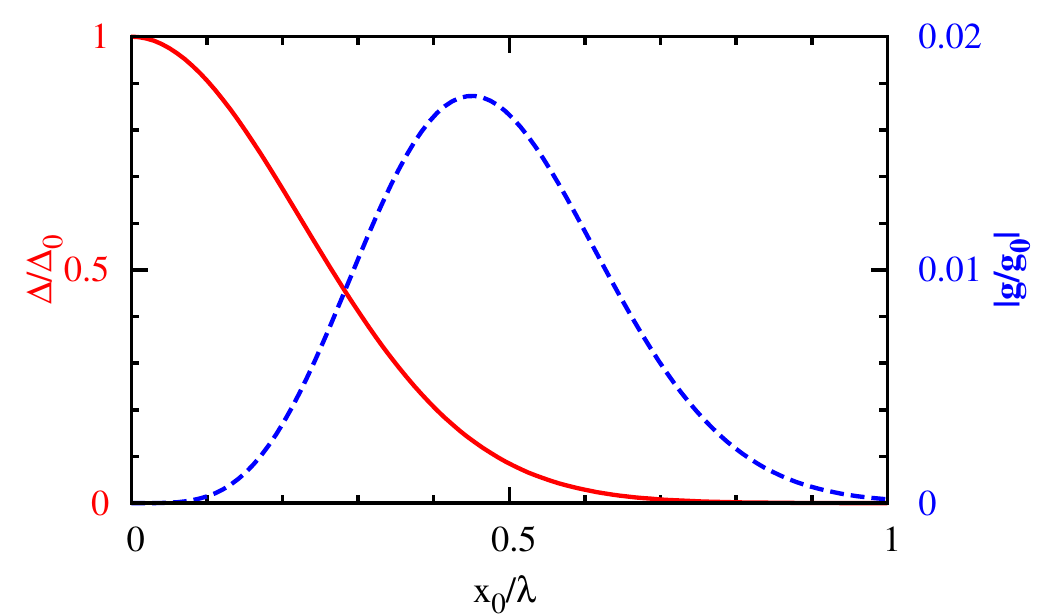}}
  \caption{(Color online) Dimensionless level spacing $\Delta/\Delta_0$ of the
    spin-orbit qubit and dimensionless spin-photon coupling $|g/g_0|$ versus the dimensionless dot size $x_0/\lambda$.}
  \label{fig2} 
\end{figure}

Note that in this device, both the level spacing $\Delta$ and
the spin-photon coupling $g$ are independent of the phase $\phi$ and proportional to the exchange coupling strength $J$. Also, both $\Delta$ and
$g$ strongly depend on the ratio between the dot size $x_0$ and
the wavelength $\lambda$ of the spiral magnetic order. In Fig.~\ref{fig2}, we plot the dependence of $\Delta/\Delta_0$
and $|g/g_0|$ on the parameter $x_0/\lambda$. One finds that when $x_0/\lambda$
is close to 1, both $\Delta$ and $|g|$ approach zero, hindering the device operation. This is because when $x_0/\lambda$ is large, 
the exchange field, oscillating with a high frequency in the scale of the dot
size, has quite small matrix elements between the $|n\pm\rangle$ and $|n^\prime\pm\rangle$ states. This
leads to a vanishing Zeeman-like splitting and spin-orbit mixing of the harmonic
oscillator states. However, in the $x_0/\lambda=0$ limit, $\Delta$ reaches its maximum while
$|g|$ again approaches zero. In fact, in this regime, with the
approximately homogeneous exchange field experienced by the quantum-dot
electron, the spin-orbit interplay becomes quite weak and
a nearly pure spin qubit with the largest Zeeman-like splitting is obtained.

\section{Modulation by the Rashba SOC}
Although the spin-orbit qubit proposed here is available without employing the
Rashba/Dresselhaus SOC, in reality the SOC may be present and
even important. 
Nonetheless, the Rashba SOC is controllable, e.g., by a
gate voltage applied to the nanowire [as illustrated by $V_g$ from the gate
electrode on top of the nanowire in Fig.~\ref{fig1}(a)]. Below we introduce
the Rashba SOC into the nanowire, supplying an effective channel to modulate the spin-orbit qubit as well as its coupling to photons.

With the Rashba SOC included, the Hamiltonian given by Eq.~(\ref{hamil}) becomes
\begin{align}
H_\alpha=\frac{p^2}{2m_e}+\frac{1}{2}m_e\omega^2x^2+\alpha p\sigma_y+{\bf
  J}(x)\cdot{\bgreek \sigma},
\label{hamil1}
\end{align}
where $\alpha$ is the Rashba SOC strength. We now apply the unitary transformation
${\tilde H}_\alpha=U_\alpha^\dagger(x)H_\alpha U_\alpha(x)$ with
$U_\alpha(x)=\exp{(-im_e\alpha x\sigma_y/\hbar)}$, and obtain \cite{zhang014433,li086805}
\begin{align}
   {\tilde H}_\alpha=\frac{p^2}{2m_e}+\frac{1}{2}m_e\omega^2x^2+{\bf J}_\alpha(x)\cdot{\bgreek \sigma}+\frac{m_e\alpha^2}{2},
\end{align}
where ${\bf J}_\alpha(x)=J(\sin(\chi q_\alpha x+\phi),0,\cos(\chi q_\alpha
x+\phi))$ with $q_\alpha=(1-\alpha/\alpha_0)q$. The Hamiltonian ${\tilde
  H}_\alpha$ has exactly the same form as in Eq.~(\ref{hamil}). Therefore, one can obtain the
low-energy eigenstates of ${\tilde H}_\alpha$ immediately based on the results
given previously. By noting that the electric dipole moment commutes with
the unitary operator $U_\alpha(x)$, one straightforwardly obtains the level spacing of the spin-orbit
qubit and the spin-photon coupling in the presence of the Rashba SOC,
\begin{align}
&{\Delta}_\alpha=\Delta_0\exp{(-\eta_\alpha^2)},\\& g_\alpha=g_0(x_0/\lambda)^3{\eta_\alpha}\exp{(-\eta_\alpha^2)},
\end{align}
with ${\eta}_\alpha=(1-\alpha/\alpha_0)\eta$.

\begin{figure}[hbt]
  {\includegraphics[width=8cm]{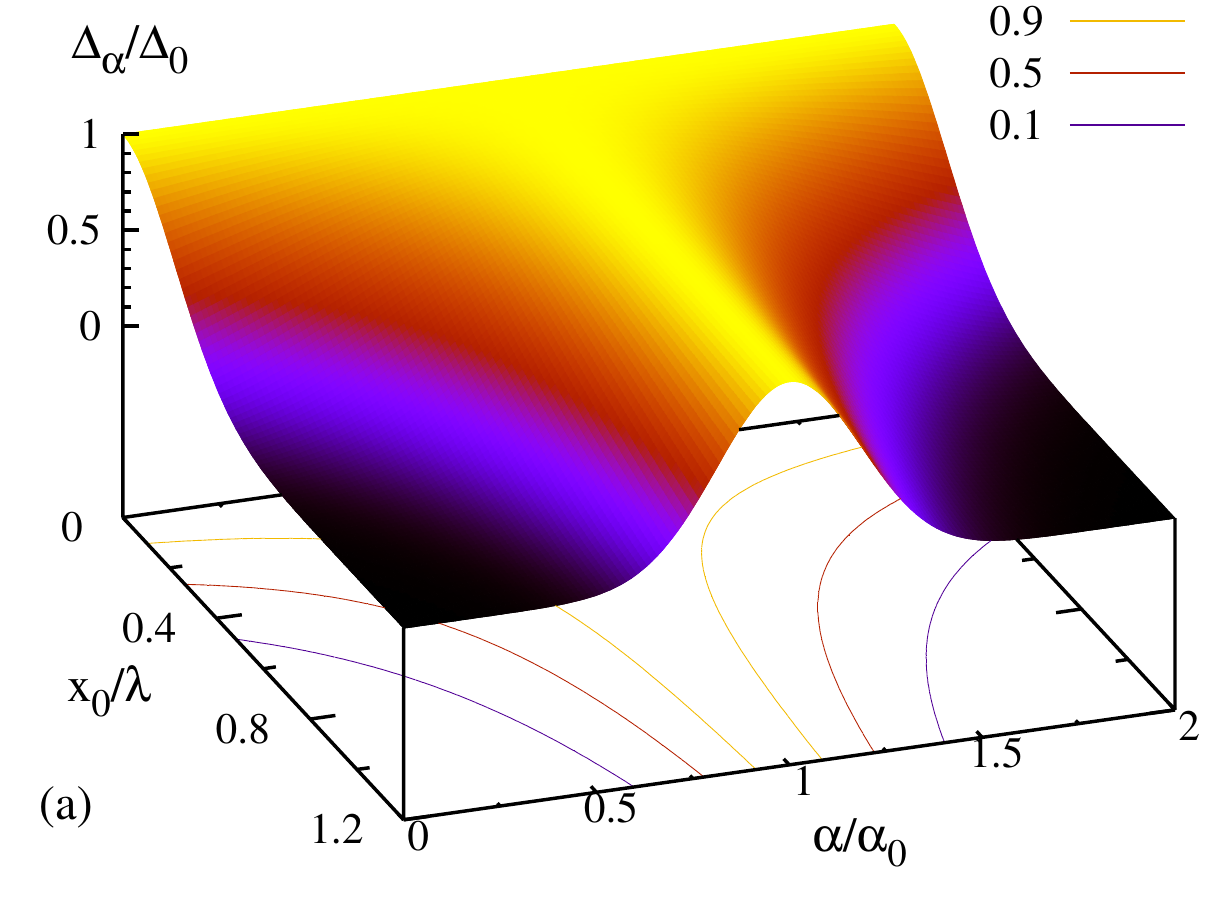}}\\{\includegraphics[width=8cm]{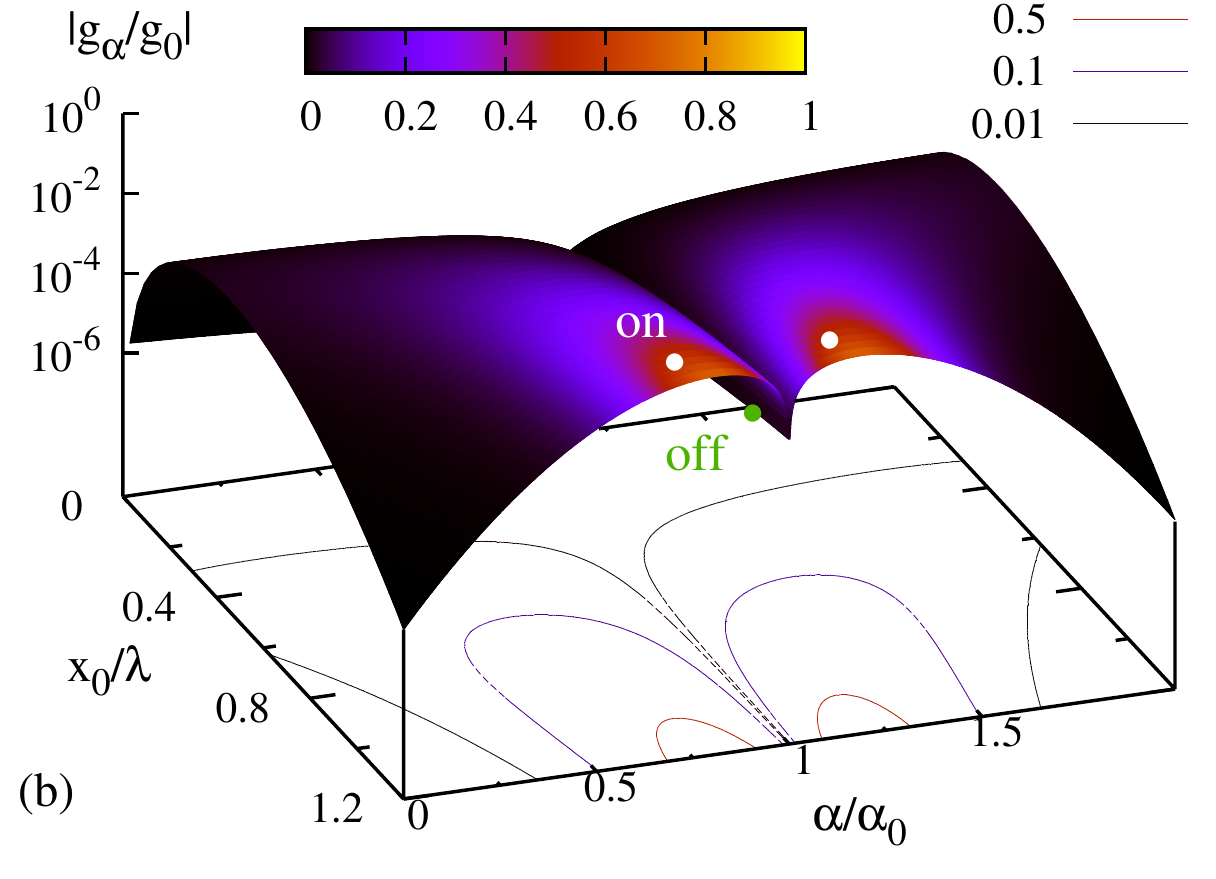}}
  \caption{(Color online) (a)~Dimensionless level spacing $\Delta_\alpha/\Delta_0$ of the
    spin-orbit qubit and (b)~dimensionless spin-photon coupling $|g_\alpha/g_0|$
    (in log-scale) versus $x_0/\lambda$ and $\alpha/\alpha_0$.}
  \label{fig3} 
\end{figure}

The above results can be understood by considering the Rashba SOC to superimpose
on the effective Rashba-like SOC from the spiral geometry, or, in other words,
to equivalently modulate the wavelength of the spiral magnetic order. This feature allows to control 
both the level spacing of the spin-orbit qubit and the spin-photon coupling by 
adjusting the Rashba SOC via the gate voltage. To show the modulation of the
Rashba SOC on the hybrid system, in Figs.~\ref{fig3}(a, b) we plot the
dimensionless level spacing, $\Delta_\alpha/\Delta_0$, and
dimensionless spin-photon coupling, $|g_\alpha/g_0|$, versus the parameters $x_0/\lambda$ and 
$\alpha/\alpha_0$. Those calculations indicate that when $x_0\sim\lambda$ and
$\alpha\sim(1\pm 0.2)\alpha_0$, the spin-orbit qubit can be effectively coupled
to photons, as indicated by the area near the ``on'' points in Fig.~\ref{fig3}(b). Moreover, by tuning $\alpha$ to
$\alpha_0$, the spin-photon coupling is completely switched off due to the
decoupling of the spin to the orbit, as indicated by the area near the``off'' point in
Fig.~\ref{fig3}(b). During this switch process, the level spacing of the
spin-orbit qubit changes by about 30\%. These features are promising
for manipulating, storing and transferring information in the hybrid quantum
systems.\cite{xiang623}


Note that for a particular Rashba SOC, its modulation depends on the spiral helicity in
the substrate, as $\alpha_0$ depends on $\chi$. This feature supplies another
control channel of the device via the gate voltage $V_c$ on the substrate, and
also in turn provides the possibility to determine the exchange coupling
strength $J$ as well as the Rashba SOC strength $\alpha$. By measuring the level spacings of the spin-orbit qubit corresponding to
opposite spiral helicities, which satisfy $\ln[\Delta_\alpha(\chi=1)/\Delta_\alpha(\chi=-1)]=2q\alpha/\omega$, one can
obtain the Rashba SOC strength $\alpha$ with the knowledge of the confining
potential of the quantum dot. Here $\alpha$ is assumed to be marginally affected
by the reversal of the spiral helicity. Further, the exchange coupling
strength $J$ is available based on the known ${\Delta}_\alpha$ and $\alpha$.

\section{Experimental realizability}
Let us now discuss the experimental realizability of the proposed spin-orbit
qubit and its coupling to the superconducting coplanar waveguide. We
consider a $\langle 110\rangle$-oriented Ge nanowire \cite{wu3165,greytak4176,niquet084301} on the surface
of the multiferroic insulator BiFeO$_3$.\cite{khomskii20,thomas423201,lebeugle024116} In the $\langle
110\rangle$-oriented Ge nanowire, the electron effective mass $m_e=0.08m_0$,
where $m_0$ is the free electron mass.\cite{niquet084301} For BiFeO$_3$, the
wavelength of the spiral magnetic order $\lambda=62$~nm,\cite{lebeugle024116,thomas423201} while the
magnon frequency is of the order of 100~GHz.\cite{rovillain975} The exchange coupling strength is set as $J=5~\mu$eV,
smaller than the estimated interface exchange coupling (16~$\mu$eV) induced by
the ferromagnetic-insulator contacts in Ref.~\onlinecite{cottet160502}. The electric field 
in the superconducting coplanar waveguide has the typical maximal strength\cite{hu035314}
$E=0.2$~V/m. With these parameters, we have $\Delta_0=(2\pi)2.4$~GHz,
$|g_0|=(\pi)0.1$~MHz, and $|\alpha_0|=7.4\times 10^4$~m/s. Moreover, even when
$x_0\sim\lambda$, the orbital splitting in the quantum dot is $\hbar\omega\sim$~0.25~meV,
still much larger than the exchange coupling strength $J$. In addition to the
availability of an effective spin-photon coupling with an efficient on/off switching, the proposed device has another
advantage. That is, in isotopically-purified $^{72}$Ge samples, the
hyperfine interaction can be markedly suppressed and hence the coherence time
of the spin-orbit qubit in the zero-temperature limit can be quite long.\cite{hu035314} This feature benefits the application of the proposed
device.

\section{Conclusion}
In conclusion, we have proposed a spin-orbit qubit based on a nanowire quantum
dot on the surface of a multiferroic insulator, and designed a hybrid quantum
circuit by integrating this spin-orbit qubit into a superconducting coplanar
waveguide.

The spiral exchange field from the magnetic moments in the multiferroic
insulator causes an inhomogeneous Zeeman-like interaction on the electron spin in the
quantum dot. This effect assists the realization of a spin-orbit qubit in the
quantum dot. In this approach, no external magnetic field is employed, benefitting the on-chip fabrication of the spin-orbit qubit in a superconducting
coplanar waveguide. Our study reveals that both the
level spacing of the spin-orbit qubit and the spin-photon coupling are
proportional to the exchange coupling strength and depend on the ratio of the dot size to the wavelength of the
spiral magnetic order. We further consider the effect of the Rashba SOC, which
is controllable by a gate voltage on the nanowire. It is found that by invoking the Rashba SOC,
one is able to obtain an effective spin-photon coupling with an efficient on/off
switching, making the device promising for applications. The proposed spin-orbit
qubit may be experimentally realizable by placing a $\langle 110\rangle$-oriented
Ge nanowire on the surface of the multiferroic insulator BiFeO$_3$.

\begin{acknowledgments}
The authors gratefully acknowledge E.~Ya.~Sherman and X.~Hu for valuable
discussions and comments. F.N. is partially supported by the ARO, RIKEN iTHES
Project, MURI Center for Dynamic Magneto-Optics, JSPS-RFBR contract No. 12-02-92100, Grant-in-Aid for
Scientific Research (S), MEXT Kakenhi on Quantum Cybernetics, and the JSPS
via its FIRST program. 
\end{acknowledgments}

\end{document}